\documentclass[12pt, letter]{emulateapj}
\usepackage{epsf}
\usepackage{graphicx}
\pdfoutput=1

\newcommand{\kband}{$K_s$-band}

\newcommand{\photoz}{photo-$z$}
\newcommand{\photozs}{photo-$z$s}

\newcommand{\Msun}{$M_{\odot}$}
\newcommand{\msun}{M_{\odot}}
\newcommand{\Sersic}{S\'{e}rsic}

\shorttitle{The Rise and Fall of Passive Disk Galaxies}
\shortauthors{Bundy et al.}

\submitted{Accepted version to appear in ApJ}

\begin{document}

\title{The Rise and Fall of Passive Disk Galaxies: Morphological Evolution
  Along the Red Sequence Revealed by COSMOS}

\author{Kevin Bundy\altaffilmark{1,13}, Claudia Scarlata\altaffilmark{2}, C.~M.~Carollo\altaffilmark{3}, Richard S.~Ellis\altaffilmark{4}, 
Niv Drory\altaffilmark{5}, Philip Hopkins\altaffilmark{1}, Mara Salvato\altaffilmark{4,10}, 
Alexie Leauthaud\altaffilmark{6,12}, Anton M.~Koekemoer\altaffilmark{7}, Norman Murray\altaffilmark{9}, Olivier Ilbert\altaffilmark{8}, 
Pascal Oesch\altaffilmark{3}, Chung-Pei Ma\altaffilmark{1}, Peter Capak\altaffilmark{2}, Lucia Pozzetti\altaffilmark{11}, Nick Scoville\altaffilmark{4}}


\altaffiltext{1}{Astronomy Department, University of California, Berkeley, CA 94705, USA}
\altaffiltext{2}{{\it Spitzer} Science Center, California Institute of Technology, 314-6, Pasadena, CA 91125, USA}
\altaffiltext{3}{Institute for Astronomy, ETH Zurich, 8092 Zurich, Switzerland}
\altaffiltext{4}{Astronomy Department, California Institute of Technology, Pasadena CA 91125, USA}
\altaffiltext{5}{Max-Planck Institut fur extraterrestrische Physik, Giessenbachstrashe, 85748 Garching, Germany}
\altaffiltext{6}{Physics Division, Lawrence Berkeley National Laboratory, 1 Cyclotron Rd., Berkeley, CA 94720, USA}
\altaffiltext{7}{Space Telescope Science Institute, Baltimore, MD 21218, USA}
\altaffiltext{8}{Laboratoire d'Astrophysique de Marseille, BP 8, Traverse du Siphon, 13376 Marseille Cedex 12, France}
\altaffiltext{9}{Canadian Institute for Theoretical Astrophysics, 60 St George Street, University of Toronto, ON M5S 3H8, Canada}
\altaffiltext{10}{Max Planck Institut fuer Plasma Physics and Excellence Cluster Boltzmannstrasse 2 Garching 85748, Germany}
\altaffiltext{11}{Osservatorio Astronomico di Bologna, Via Ranzani, 1, I-40127 Bologna, Italy}
\altaffiltext{12}{Berkeley Center for Cosmological Physics, University of California, Berkeley, CA 94720, USA}
\altaffiltext{13}{Hubble Fellow}

\begin{abstract}

  The increasing abundance of passive ``red-sequence'' galaxies since $z \sim 1$--2 is mirrored by a
  coincident rise in the number of galaxies with spheroidal morphologies.  In this paper, however,
  we show that in detail the correspondence between galaxy morphology and color is not perfect,
  providing insight into the physical origin of this evolution.  Using the COSMOS survey, we study a
  significant population of red-sequence galaxies with disk-like morphologies.  These passive disks
  typically have Sa--Sb morphological types with large bulges, but they are not confined to dense
  environments.  They represent nearly one-half of all red-sequence galaxies and dominate at lower
  masses ($\lesssim 10^{10} \msun$) where they are increasingly disk-dominated.  As a function of
  time, the abundance of passive disks with $M_* \lesssim 10^{11} \msun$ increases, but not as fast
  as red-sequence spheroidals in the same mass range.  At higher mass, the passive disk population
  has declined since $z \sim 1$, likely because they transform into spheroidals.  Based on these
  trends, we estimate that as much as 60\% of galaxies transitioning onto the red sequence evolve
  through a passive disk phase.  The origin of passive disks therefore has broad implications for
  our understanding of how star formation shuts down.  Because passive disks tend to be more
  bulge-dominated than their star-forming counterparts, a simple fading of blue disks does not fully
  explain their origin.  We explore the strengths and weaknesses of several more sophisticated
  explanations, including environmental effects, internal stabilization, and disk regrowth during
  gas-rich mergers.  While previous work has sought to explain color and morphological
  transformations with a single process, these observations open the way to new insight by
  highlighting the fact that galaxy evolution may actually proceed through several separate stages.

\end{abstract}


\section{Introduction}\label{intro}

It has long been known that nearby galaxies roughly fall into two categories: 1) systems with
prominent disks often exhibiting spiral structure and ongoing star formation, and 2) galaxies
dominated by a spheroidal morphology---the most extreme being ellipticals---and harboring little if
any star formation.  With the advent of large galaxy surveys like Sloan Digital Sky Survey (SDSS) it had been possible to
quantify the bimodal nature of the local population both in terms of morphology and star formation
as a function of what appears to be the best observationally accessible variable for separating the
two populations, the galaxy's stellar mass \citep[$M_*$;][]{bell03a, kauffmann03, baldry04, balogh04}.
The bimodal nature of the galaxy population is apparent to $z \sim 1$ \citep{bell04} and beyond
\citep[e.g.,][]{van-dokkum06}, but it is clear that the fraction of galaxies populating either side
must evolve with time.  \citet{brinchmann00} presented evidence for an increase since $z \sim 1$ in
the global mass density of morphological spheroidals at the expense of star-forming irregulars, and
\citet{cowie96} showed that increasingly $K$-band luminous and therefore more massive galaxies exhibit
higher specific star formation rates (sSFRs) at early times.

A number of more recent studies have confirmed and extended these early results, using mass and
luminosity functions to trace the evolution since $z \sim 2$ of the two populations as partitioned
by the amount of star formation or the rest-frame color \citep{juneau05, bundy06, borch06,
  willmer06, brown07, ilbert10, mobasher09} and as a function of morphological type \citep{bundy05, pannella06,
  franceschini06, scarlata07, abraham07, van-der-wel07, ilbert10, pozzetti09, oesch10}.  The importance of
environmental density has also been studied \citep{bundy06, cucciati06, cooper08, capak07, van-der-wel08a, 
  pannella09, tasca09, bolzonella09}.  The emerging picture is one in which, beginning at the
highest masses, galaxies transform from blue, star-forming, disk-dominated systems into red,
bulge-dominated spheroidals in which star formation has been quenched.  As time proceeds, this
transformation process works its way down the mass function, driving the apparent downward shift in
the bimodality mass scale.

Understanding the physical driver of this evolution has been a key challenge, leading to a search
for a process capable of simultaneously shutting down star formation and turning disks into
spheroidals.  Morphological evolution has often been attributed to the ``merger hypothesis''
\citep{toomre77}, in which galaxy mergers convert rotation-supported disks into pressure-supported
spheroids.  In a series of papers, Hopkins and collaborators demonstrate the feasibility of this
idea through detailed simulations and link mergers with an explosive quasar or starburst phase that
may heat or drive out cold gas in the spheroidal merger remnant, thereby quenching star formation
\citep[see][and references therein]{hopkins08}.  The idea that mergers quench star formation, at
least in halos above a critical mass, $M_{\rm crit} \sim 10^{12} \msun$ \citep[where the cooling
time is longer than the dynamical time, see][]{dekel06a}, has also been implemented in semi-analytic
models, in which supposedly spheroidal merger remnants are kept ``red and dead'' by so-called
``radio mode'' active galactic nucleus (AGN) feedback \citep{croton06, bower06, cattaneo06, somerville08}.  \citet{birnboim07}
argue against the need for mergers and AGN feedback, relying instead on the interplay between shock
heating and cold flows in halos near $M_{\rm crit}$ \citep[see also][]{dekel09}.  Semi-analytic models
based on this framework have needed to include some additional quenching tied to bulge growth after galaxy
mergers to match the observed fraction of passive galaxies \citep{cattaneo08}.

One way to distinguish between these scenarios and gain insight into the transformation mechanisms
at work is to test the link between quenching and morphological evolution by studying those galaxies
whose morphological type places them on one side of the bimodality but whose star formation
identifies them with the other.  Star-forming blue ellipticals are one example
\citep[e.g.,][]{schawinski09, kannappan09, ferreras09, brand09} and are understood to be a
combination of recent arrivals to the red sequence \citep{ruhland09}, bulge-dominated systems
experiencing triggered star formation after minor mergers \citep{kaviraj09} or, at low masses, the
formation of new disks in gas-rich environments \citep{kannappan09}.  The other possibility, namely
quenched or passive disk galaxies, is more challenging to explain because the shutdown of star
formation must leave the disk in place.  Yet examples in the local universe do exist.  Lenticular or
S0 galaxies are one example, although they typically live in galaxy clusters where interactions with
the intercluster medium \citep[e.g., ram pressure stripping, see][]{moran07a} can remove the gas
supply and shut down star formation.  However, S0's have properties that do not depend on cluster
density and a non-negligible fraction ($\sim$8\%) of nearby S0's are isolated, suggesting that not
all S0's form as a result of living in dense environments \citep{van-den-bergh09}.  At slightly
higher redshifts, $z \approx 0.1$, axis ratio distributions and Galaxy Zoo (GZ) morphologies have
demonstrated an unmistakable population of red, disk-dominated galaxies that appear to live in the
field and slightly overdense environments, but are not found solely in dense regions
\citep{bamford09, masters10, van-der-wel09}.  Finally, using the {\it Hubble Space Telescope} ({\it
  HST}) data in the Great Observatories Origins Deep Survey (GOODS) fields at $z \sim 1$,
\citet{bundy06} compared color and morphologically classified mass functions to argue that quenching
appeared to precede spheroid formation, finding hints of a red disk population at $z \sim 1$
\citep[see also][]{pozzetti09}.  Some examples of passive disks have even been found at $z \sim 2$
\citep{stockton08}.

The goal of this paper is to utilize the unprecedented size of the COSMOS field with its 1.64 deg$^2$
of {\it HST} imaging to revisit the question of red, passive disks in field environments and quantify
their origin and evolution since $z \sim 1$.  We argue that the behavior of this population presents
a challenge to models that attribute morphological and color evolution to a single event, and may
shed light on mechanisms that affect the galaxy population more broadly.  This work follows on mass
function studies of the COSMOS field as a function of color and morphological type by
\citet{ilbert10}, \citet{pozzetti09}, \citet{oesch10}, and \citet{drory09}.  \citet{ilbert10}
specifically showed a broad connection in the behavior of morphologically early-type galaxies and
those with passive stellar populations, but also presented evidence for an increasing
fraction at high redshift of non-spheroidal morphologies among quenched systems---this is the
starting point for the present study.

Throughout this paper, we use the AB magnitude system and adopt a
standard cosmology with $H_0$=70 $h_{70}$ km s$^{-1}$ Mpc$^{-1}$,
$\Omega_M$=0.3 and $\Omega_{\Lambda}$=0.7. 

\section{Observations}\label{data}

\subsection{The COSMOS Field}

The COSMOS field ($10^h00^m28.6^s, +02^\circ12$\arcmin21.0\arcsec) is the largest contiguous area
(1.64 deg$^2$) imaged by the {\it HST} and is almost 20 times the size of
the two GOODS fields combined.  As such, it marks an unprecedented step forward in studies of the
morphologies of distant ($z \sim 1$) galaxies at high resolution.  Specifically for this work,
COSMOS provides such large samples that we can statistically track evolving sub-samples of field
galaxies defined by both rest-frame color {\em as well as} morphology as a function of $M_*$ and
redshift from $z \sim 1$ to $z \sim 0.1$.

A large team of scientists has assembled to collaborate on a wide range of topics using observations
from the COSMOS field \citep[for a review, see][]{scoville07a}.  The primary data sets of interest
for this work are the $i_{814W}$ observations from the {\it HST} Advanced Camera for Surveys
\citep[ACS;][]{scoville07, koekemoer07}, which provide morphological information, and the
ground-based optical to near-IR photometry presented in detail in \citet{capak07a} and updated in
\citet{ilbert09}.  As we discuss further below, the ground-based catalog provides the basis for high
precision photometric redshifts and a set of rest-frame colors that separate truly passive stellar
populations from those that are actively star-forming but dusty.

We briefly review the relevant components of the ground-based catalog, referring to \citet{capak07a}
and \citet{ilbert09} for details.  The catalog includes $u^*$- and $i^*$-band imaging from
Mega-Prime \citep{aune03, boulade03} and was cross-referenced with \kband\ observations
\citep{mccracken10} from the Wide-field InfraRed Camera \citep[WIRCAM,][]{puget04}, both of which
are instruments on the Canada-France-Hawaii Telescope (CFHT).  The majority of photometric bands in
the catalog were observed with the Subaru Telescope equipped with Suprime-Cam \citep{komiyama03}
(filters $B_J, V_J, g^+, r^+, i^+, z^+$).  The catalog was selected based on detections in the
Subaru $i^+$-band and the seeing of the final mosaics is better than 1\farcs5 in all
cases.  \citet{capak07a} point-spread function (PSF) match all data, optimizing for photometry
performed in 3\arcsec\ diameter apertures.  All of the optical bands have a 5$\sigma$ point-source
(3\arcsec\ diameter aperture) depth of at least 25th magnitude (AB) with the \kband\ limited to $K_s
< 24$.  Given these depths, the final sample completeness is determined mostly by the ACS data
because morphological classification begins to degrade at magnitudes fainter than $i_{814W} \approx
24$.  The {\it HST} ACS imaging is described in \citet{scoville07} and \citet{koekemoer07}, and is used only for
morphological classification in this work.  We use the Zurich Estimator of Structural Types (ZEST)
morphology catalog \citep{scarlata07a}---discussed further in Section \ref{morph}---which is based on
the ACS source catalog presented in \citet{leauthaud07} who estimate the ACS depth at $i_{814W}
= 26.1$ for photometry of a 1\arcsec\ galaxy.

\subsection{The Primary Sample}

\begin{deluxetable*}{lcccccc}
\tablecaption{Sample Statistics}
\tabletypesize{\footnotesize}
\tablewidth{0pt}
\tablecolumns{7}
\tablehead{
\colhead{$z$} & \colhead{All} & \colhead{Passive} & \colhead{Star-forming} & \colhead{Passive Early Disk} & \colhead{Passive Late Disk} & \colhead{Passive Extreme Disk} \\
}

\startdata

$0.2< z <0.4$ & 17210 & 2402 & 14808 & 755 & 380 & 30 \\
$0.4< z <0.6$ & 16324 & 1744 & 14580 & 523 & 171 & 17 \\
$0.6< z <0.8$ & 22253 & 2821 & 19432 & 883 & 261 & 36 \\
$0.8< z <1.0$ & 25581 & 4075 & 21506 & 1282 & 356 & 81 \\
$1.0< z <1.1$ & 6014 & 480 & 5534 & 160 & 42 & 5 \\
$1.1< z <1.2$ & 14860 & 517 & 14343 & 165 & 28 & 5 \\

\enddata
\label{table:sample}
\end{deluxetable*}

In the analysis that follows, our primary sample is based on the $i$-selected catalog from
\citet{capak07a}.  We impose magnitude limits of $K_s < 24$ and $i < 26.5$ and divide our sample
into six redshift bins: [0.2--0.4), [0.4--0.6), [0.6--0.8), [0.8--1.0), [1.0-1.1), [1.1--1.2).  Given
the magnitude cuts, we estimate the mass completeness of each redshift bin by considering the
observed magnitude of a maximal $M_*/L$ stellar population model with solar metallicity, no dust,
and a $\tau = 0.5$ Gyr burst of star formation that occurred at $z_{\rm form} = 5$.  This exercise
provides limits that roughly match the 80\% completeness limits determined when deeper samples are available
\citep{bundy06}, and in this case gives values of 8.8, 9.3, 9.7, 10.0, 10.1, and 10.2 in units of
$\log M_* / \msun$.

In practice, morphological classification is not possible for every source down to the $K_s$- and
$i$-band limits.  This means that the effective $M_*$ above which morphological samples are complete
is higher than the full sample.  These depend on the nature of the source, however, but can be
easily defined with respect to the deeper $K_s < 24$ and $i < 26.5$ sample.  We consider the
morphological sample incomplete when the fraction of galaxies with a successful ZEST morphological
classification (see Section \ref{morph}) falls below 50\% of the primary sample.  We also limit the sample
to ensure the robustness of ZEST classifications using ellipticity distributions tests (see
Section \ref{verification}).  For each of the redshift bins defined above, we adopt mass limits of: 9.7,
9.9, 10, 10.4, 10.5, 10.8 in units of $\log M_* / \msun$.  After removing stars and bright AGN, the
number of galaxies in the primary sample in each redshift bin is listed in Table \ref{table:sample}.

\subsubsection{Cosmic Variance}\label{cosmic_variance}

While population studies using COSMOS data largely agree with previous work based on smaller fields
(see Section \ref{intro}), the sample or cosmic variance even in surveys covering a few square degrees
represents a significant uncertainty that must be considered when interpreting our results
\citep[see][]{meneux09}.  The largest effect is on the overall normalization of the total galaxy
abundance \citep[e.g.,][]{bundy06} although fractional abundance values are also affected.
\citet{stringer09} use the variance gleaned from multiple mock light cones based on the Millennium
Simulation to estimate the cosmic variance signal for a number of surveys, including COSMOS.  From
one redshift interval to another, they find typical abundance uncertainties of 0.2 dex.  This
estimate is consistent with the results in this work.

\section{Methods}\label{properties}

\subsection{ Estimating Photometric Redshifts}\label{redshifts}

While a spectroscopic survey, $z$COSMOS, is ongoing in the COSMOS field \citep{lilly07}, the large
samples required for this work necessitate the use of photometric redshifts.  We use the catalog of
\photozs\ published by \citet{ilbert09} which are based on 30 bands of imaging in the COSMOS field.
In addition to the observations mentioned in Section \ref{data}, these include UV bands from the {\it
  Galaxy Evolution Explorer} ({\it GALEX}), four mid-IR bands (3.6--8 $\mu$m) from the Infrared Array Camera
(IRAC) on the {\it Spitzer Space Telescope}, $J$-band imaging from the Wide Field IR Camera (WFCAM)
on the UK IR Telescope (UKIRT), and an additional 12 medium-band and two narrowband filters observed
with Suprime-Cam on Subaru.  This extended data set is described in detail by P.~Capak et al.~(in preparation).

Redshift estimates were obtained with the {\it Le
  Phare}\footnote{www.oamp.fr/people/arnouts/LE\_PHARE.html} code using templates drawn from
\citet{polletta07}.  \citet{ilbert09} develop a novel technique that accounts for the contribution
of emission line flux (especially important in the narrow and medium filter bands) to the spectral
energy distributions (SEDs) of star-forming
galaxies.  This correction improves the \photoz\ precision by a factor of $\sim$2.  When comparing
to $z$COSMOS spectroscopic redshifts with $17.5 \leq i^+ \leq 22.5$ \citet{ilbert09} find a
precision of $\sigma_{\Delta z} / (1+z) \approx 0.01$ after removing an outlier fraction that
accounts for 0.6\% of the comparison sample.  They also compare to spec-$z$s obtained with Keck
DEIMOS \citep{faber03} for a sample of very faint and 24 $\mu$m-selected sources.  At $i^+ \approx
25.5$ the uncertainties are estimated at $\sigma_{\Delta z} / (1+z) \approx 0.06$.

\subsection{Estimating Stellar Masses}

We use the Bayesian code described in \citet{bundy06} to estimate the current mass in stars of the
galaxies in our sample.  The observed SED of each galaxy is compared to a grid of 13440
models from the \citet{bruzual03} (BC03) population synthesis code that spans a range of metallicities, star
formation histories (parameterized as exponentials), ages, and dust content.  No bursts are included
in our models and only those models with ages (roughly) less than the cosmic age at each redshift
are considered.  We use a Chabrier initial mass function \citep[IMF;][]{chabrier03} and assume a Hubble constant of 70 km
s$^{-1}$ Mpc$^{-1}$.

At each grid point, the \kband\ $M_*/L_K$ ratios, inferred $M_*$, and probability that the model
matches the observed SED is stored.  This probability is marginalized over the grid, giving the
stellar mass probability distribution, the median of which is taken as the final estimate of $M_*$.
The width of the distribution provides the uncertainty which is typically 0.1--0.2 dex.  This is
added in quadrature to the \kband\ magnitude uncertainty to determine the final error on $M_*$.
Stellar mass estimates for galaxies with \photozs\ also suffer from the uncertainty in luminosity
distance introduced by the \photoz\ error and the possibility of catastrophically wrong redshift
information \citep{bundy05, kitzbichler07}.

More broadly, any stellar mass estimate suffers potential systematic errors\footnote{These are not
  included in our final error estimates.} from uncertainties inherent in stellar population modeling and
various required assumptions, such as the form of the IMF.  Several papers have stressed the
importance of treating thermally pulsating asymptotic giant branch stars \citep[TP-AGBs,
e.g.,][]{maraston06} an element that is missing in the \citet{bruzual03} models.  The recent and
thorough investigation of population synthesis modeling by \citet{conroy09}, however, argues that
$M/L$ ratios estimated from \citet{bruzual03} are largely resistant to the uncertain contribution
from TP-AGBs as well as other limitations of current models.  A further test is provided in
\citet{drory09} who compare mass functions computed using BC03 and BC07 stellar population codes and
conclude that $\sim$20\% differences in abundances can be introduced, especially for star-forming
galaxies.  It is also important to recognize that $M_*$ estimates may be affected by unrecognized
systematic uncertainties at the 0.1--0.2 dex level.

\subsection{Excluding Dusty Red-sequence Galaxies}\label{bicolor}

Because our goal in this paper is to study disk galaxies that have stopped forming stars, in this
section we discuss the rest-frame galaxy colors we use to identify passive stellar populations.
Previous work has often relied on a single rest-frame UV-optical color to select passive galaxies,
but this technique leads to possible contamination from dusty star-formers \citep[e.g.,][]{yan06}.
In this work, we make use of color-color diagrams that include rest-frame UV, optical, and near-IR
colors, inspired by recent work showing how this approach separates dusty star-formers from truly
passive populations \citep[e.g.,][]{pozzetti00, labbe05, wuyts07, williams09}.

\begin{figure*}
\centerline{\includegraphics[scale=0.8]{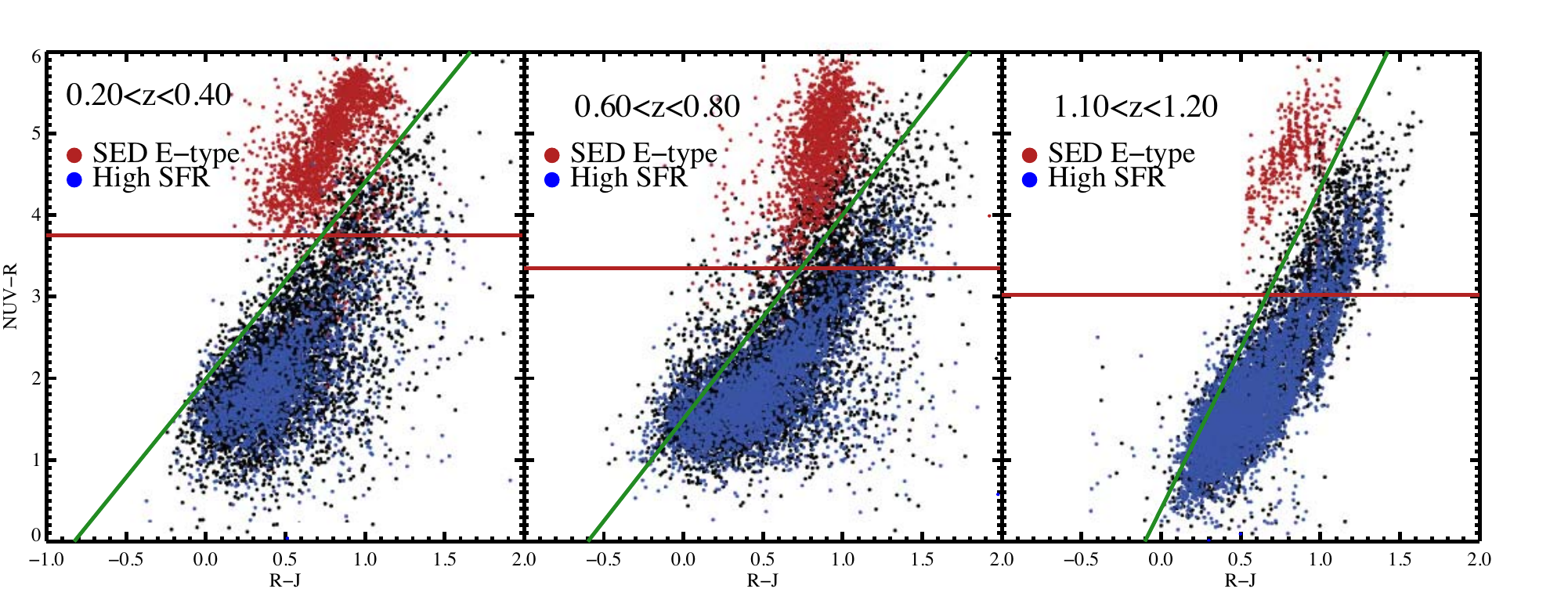}}
\caption{Rest-frame (NUV$-R$) vs.~($R-J$) color--color diagrams in three
  redshift bins selected to represent the full $z$-range of our sample
  ($0.2 < z < 1.2$).  As shown by
  \citet{williams09}, the addition of a near-IR filter helps break the
  degeneracy between dusty star-forming galaxies (red in NUV$-R$ and red
  in $R-J$) and truly passive populations, which form a tight locus
  that is red in NUV$-R$ but blue in $R-J$.  The lines indicate the
  adopted color cuts.  Red and blue colors indicate early and highly star-forming SED
  types, respectively, as identified by the \photoz\ fitting code of
  \citet{ilbert09}.  \label{fig:bicolor}}
\end{figure*}

The utility of the rest-frame UV-optical-near-IR ``bicolor'' diagram for the COSMOS data set is
illustrated in Figure \ref{fig:bicolor} where we show interpolated rest-frame colors in various
redshift bins using the following filter bands: the GALEX near-UV band (NUV), the Suprime-Cam
$r^+$ band which we label as $R$, and a generic $J$-band filter\footnote{As this work was completed,
  $J$-band observations of the COSMOS field were obtained using the WFCAM on UKIRT which has a
  different filter response function than the generic $J$-band filter used here.}.  Rest-frame
NUV- and $R$-band absolute magnitudes were estimated by \citet{ilbert09} during the SED fitting
necessary for measuring \photozs.  To derive the rest-frame $J$-band absolute magnitudes needed for
the NUV-$R$-$J$ diagram, we used the {\tt Kcorrect} software package \citep{blanton07} and the
full suite of photometry described in Section \ref{data}.  The broad wavelength coverage of COSMOS ensures
that the flux corresponding to a given rest-frame filter is well sampled.  A similar diagram was
discussed in \citet{ilbert10}.

The upper-left portion of each redshift panel in Figure \ref{fig:bicolor} exhibits a relatively
tight clump of galaxies that we associate with passive stellar populations.  A sequence of
star-forming galaxies is evident below and to the right of the clump in each panel.  As a test, we
have colored in red those galaxies with SEDs best fit by early ``spectral'' type templates, as
determined by the \photoz\ code \citep[this category includes types 1--8 in the COSMOS \photoz\
fits, see][]{ilbert09}.  Highly star-forming templates are shown in blue.  At most redshifts, the
early-type SED designation overlaps well with the associated passive clump.  Note that the location
and orientation of the passive clump and star-forming sequence evolves somewhat given our choice of
filters.  In what follows, we will isolate passive clump galaxies (also referred to hereafter as the
``red sequence'') using a joint color cut that evolves accordingly and is indicated by the two lines
in Figure \ref{fig:bicolor}.  Recognizing that such color cuts are somewhat arbitrary, our goal is
to evenly split the now meandering ``green valley'' that separates the two populations.  We specify
our thresholds in terms of the given filters because they correspond well to our observations and do
a good job of making the passive clump distinct, although we acknowledge that a more general
treatment would be valuable in future work.  Still, we emphasize that adjusting the details of the
cuts defined below has little effect on the analysis that follows.

The horizontal NUV$-R$ color cut is tuned to the usual luminosity dependent bimodality
\citep[e.g.,][]{willmer06} in this color and is defined as
follows:

\begin{equation}
{\rm NUV} - R > 4.2(1+z)^{-0.43} - 0.2(M_K + 20),
\end{equation}

\noindent where $M_K$ is the rest-frame absolute  \kband\ magnitude.  Passive galaxies must also
satisfy the following diagonal cut in Figure \ref{fig:bicolor}:

\begin{equation}
{\rm NUV} - R > C_1(z) + C_2(z)(R-J),
\end{equation}

\noindent where the constants, $C_1$ and $C_2$, have been chosen by inspection in redshift bins of
width 0.2 with central values of $z = [0.30, 0.50, 0.70, 0.85, 0.95, 1.10]$.  They are given by
$C_1(z) = [4.4, 4.2, 4.0, 3.9, 3.8, 4.2]$ and $C_2(z) = [2.41, 2.41, 2.5, 2.6, 3.0, 3.7]$.  At $z
\lesssim 0.6$ the combination of the color cuts above select 10\%--20\% more passive galaxies than
those with early-type SED templates, many of them scattered between star-forming and passive
sequences.  At higher redshifts, however, the SED method appears to miss genuine galaxies in the
passive clump, as many as 20\% at $z \approx 1$.  We note that our conclusions regarding passive
disk galaxies are unchanged if we use the SED fits to define the passive
population.  

Finally, we have checked for obscured star formation among the passive population as indicated by
MIPS 24$\micron$ emission.  We note that a significant fraction of 24$\micron$ emission, especially
among the passive population, might arise from AGN \citep[e.g.,][]{fiore08}.  In addition, among
star-forming galaxies, the implied SFR from 24$\micron$ emission must be added to that from the UV
flux which will dominate the output luminosity from star formation in our sample
\citep[e.g.,][]{noeske07}.  As a result, our constraints on hidden star formation in passive
galaxies compared to star-formers are upper limits.  Using the morphological classifications defined
below (grouping ``early'' and ``late'' disks together), our first test compares 3$\sigma$ MIPS
detection rates of passive versus star-forming disk galaxies.  For $z > 1$, no passive disks are
detected with MIPS while the detection rate among star-forming disks is roughly 10\%.  In our lowest
redshift bin, $0.2 < z < 0.4$, 1\% of passive disks are detected compared to 10\% of star-formers.
At redshifts in between, the highest detection rates of passive disks occur at $0.4 < z < 0.6$,
where 5\% are detected compared to 8\% for star-forming disks.  Considering the full redshift range, these
numbers confirm that the color-color methodology described above does not lead to significant
contamination from hidden star formation.  However, given the small numbers of MIPS detections, it is not
possible to compare 24$\micron$ flux as a function of $M_*$ or put constraints on the relative
amount of SFR hidden among passive population but below the MIPS detection limits.

For this, we perform a stacking analysis of the MIPS data in bins of stellar mass (of width 0.5 dex)
and in the redshift intervals used throughout (the minimum number of galaxies per bin is roughly
20).  Using \citep{rieke09}, we estimate mean sSFR given the stacked
24$\micron$ emission.  Across our $\sim$30 stacked bins, the average passive sSFR is roughly an
order of magnitude less than the star-forming sSFR.  In two bins this difference approaches a factor
of 2, and in several other bins, there is no stacked MIPS detection for passive galaxies.  We note
that the IR/SFR calibrations become uncertain at low 24$\micron$ luminosities below $\log
L_{24}/L_{\odot} < 8.5$ and several of our bins reach this limit.  With this caveat and again ignoring
contributions from AGN, our stacking analysis provides a rough upper limit to the possible amount of
hidden star formation (SF) in the average passive galaxy that corresponds to growth rates of 3\%--5\% in $M_*$ per Gyr.

\subsection{Morphological Classification}\label{morph}

Understanding the morphological makeup of the galaxy population is crucial to our goals in this
paper.  Given the large samples (typically more than $10^5$ galaxies) made possible by the COSMOS
data, automated morphological classifiers figure prominently.  It is well known, however, that
techniques such as CAS \citep[e.g.,][]{conselice03} or Gini/M20 \citep[see][]{abraham07} often disagree
when compared to visual morphologies and can differ from each other.  Indeed, \citet{ilbert10}
use two different classifiers---the Gini/M20 method described in \citet{abraham07} and a CAS-like
technique used in \citet{cassata07}---to gauge the systematic uncertainty of broad categories such as
``spheroidal'' or ``disk-like.''  Their analysis of the evolving distribution of morphologies in
COSMOS from $z \sim 1$ serves as the basis for this work.  We should also note that
``wiki-morphologies'' of the kind pioneered by GZ \citep{lintott08} are an alternative way to
obtain visual morphologies for large samples.

In this work, we are specifically interested in disk components living in passive galaxies.  As
such, the exact definitions of morphological type are less important than confidence in the ability
of the chosen classifier to find disk-like structure even if embedded in bulge-dominated galaxies
and, ideally, to parametrize the prominence of disks compared to an accompanying bulge component.
The ZEST automated classifier satisfies these requirements and has also been
trained to match visual classifications of local surveys.  Our strategy is to adopt the ZEST
classification in the COSMOS sample and test it both through visual inspection of COSMOS galaxies
and an independent check of our results using visual morphologies determined in GOODS-N and
published in \citet{bundy05}.

\subsubsection{The ZEST Automated Classifier}

ZEST combines the power of a principle component (PC) analysis of nonparametric measures of galaxy
structure with information from a parametric fit.  Details are presented in \citet{scarlata07a} but
we review the salient features here.  The light concentration, asymmetry, Gini coefficient,
second-order moment of the brightest 20\% of the flux, $M_{20}$, and the ellipticity are input to a
PC analysis which delivers eigenvectors (linear combinations of the input parameters) that maximize
the observed sample variation, thereby allowing for a compression of the original parameter space.
Only the first three PCs, PC$_1$, PC$_2$, and PC$_3$, are retained for further analysis
because they contain the bulk of the sample variance.  The PC$_1$--PC$_2$--PC$_3$ space defined by a
training sample of 56,000 $i_{814W} < 24$ galaxies was divided into a three-dimensional grid.  Every galaxy living
in a given unit cube was visually inspected so that a broad morphological type could be assigned to
that cube.  A value of $T_{\rm ZEST} = 1$ corresponds to ``elliptical,'' $T_{\rm ZEST} = 2$ to
``disk,'' and $T_{\rm ZEST} = 3$ to ``irregular.''  In the case of $T_{\rm ZEST} = 2$ (disk) cubes,
the average value $\langle n \rangle$ from single-S\'{e}rsic fits for galaxies with $i_{\rm 814W} <
22.5$ from \citet{sargent07} was used to establish a ``bulge'' parameter, $B_{\rm ZEST}$, for each
cube which was then applied to fainter magnitudes.  Higher values of $B_{\rm ZEST}$ indicate smaller
bulges and larger disks, with $B_{\rm ZEST} = [0,1,2,3]$ corresponding to S\'{e}rsic $n$ ranges of $
\langle n \rangle \geq 2.5, 1.25 \leq \langle n \rangle \leq 2.5, 0.75 \leq \langle n \rangle \leq
1.25, 0 \leq \langle n \rangle \leq 0.75$.  In the analysis that follows, we consider three ZEST
morphological types:

\begin{itemize}
\item ``Spheroidal'' includes $T_{\rm ZEST} = 1$ ellipticals as well as
highly bulge-dominated S0-like systems with $T_{\rm ZEST} = 2$ and $B_{\rm ZEST} = 0$ ($\langle n \rangle \geq 2.5$).

\item ``Early disks'' are likely to host significant bulges and have $T_{\rm ZEST} = 2$ and $B_{\rm
    ZEST} = 1$ ($1.25 \leq \langle n \rangle < 2.5$).

\item ``Late disks'' likely have smaller bulges with $T_{\rm ZEST} = 2$ and $B_{\rm ZEST} = 2$ ($0.75
  \leq \langle n \rangle < 1.25$).

\item ``Extreme disks'' have little to no bulge component with $T_{\rm ZEST} = 2$ and $B_{\rm ZEST}
  = 3$ ($0 \leq \langle n \rangle < 0.75$).

\end{itemize}

The ZEST morphologies were previously tested using a sample of 80 visually classified $z=0$ galaxies
drawn from the RC3 catalog \citep{de-vaucouleurs91, frei96}.  As shown in Figure 6 of
\citet{scarlata07a}, the ZEST classifications agree well with the published morphologies.  Of
relevance to the current work, galaxies with $T_{\rm ZEST} = 2$ (disks) and $B_{\rm ZEST} > 0$ are
always associated with types later than S0 with no contamination from ellipticals.  Of disks with
$B_{\rm ZEST} = 1$, 43\% have published types between S0 and Sab and 57\% have types Sb-Scd.  Of
disks with $B_{\rm ZEST} = 2$, 5\% are classified as S0-Sab, 45\% as Sb-Scd, and 50\% as Sd type and
later.

At faint magnitudes the ZEST morphologies begin to break down, as discussed further in
\citet{scarlata07a}.  This can be quantified by degrading the signal-to-noise ratio (S/N) of bright galaxies for which the
morphology is well determined.  Down to $i_{814W} = 22.5$ fewer than 10\% experience any change in
either $T_{\rm ZEST}$ or $B_{\rm ZEST}$ and those that do mostly lie near classification boundaries.
Even when degraded to $i_{814W} = 24$, the effective limit for morphological classification, the
fraction identified with their original morphological type is 65\%.  Because we will be interested
in trends between morphology, mass, and redshift, it is important to note that at faint magnitudes,
the ZEST morphologies are biased to later types.  \citet{scarlata07a} estimate that at most 30\% of
early-type galaxies could be misclassified as disk or irregular type at the faintest magnitudes.  We
will show that this potential bias is smaller than the detected amount by which disks outnumber
spheroidals at the faint ends of our sample.  Furthermore, imposing brighter magnitude limits (e.g.,
$i_{814W} < 22.5$) to reduce the potential for poor classifications does not alter our basic
conclusions.

\subsection{Estimating Comoving Number Densities}\label{method_MF}

We compute mass functions in this work using the $V_{\rm max}$ technique \citep{schmidt68}.  We weight
galaxies by the maximum volume in which they would be detected within the \kband\ and $i^+$-band
limits in a given redshift interval.  In practice, our magnitude limits ($K < 24$ and $i^+ < 26.5$)
impose restrictions only in the highest redshift bins ($z \gtrsim 1$) and at masses below the limit
where morphological classification becomes difficult.  For each host galaxy $i$ in the redshift
interval $j$, the value of $V^i_{\rm max}$ is given by the minimum redshift at which the galaxy would
drop out of the sample

\begin{equation}
V^i_{\rm max} = \int_{z_{\rm low}}^{z_{\rm high}} d \Omega \frac {dV}{dz} dz,
\end{equation}

\noindent where $d \Omega$ is the solid angle subtended by the survey area, and $dV/dz$ is the comoving
volume element.  The redshift limits are given as,

\begin{equation}
z_{\rm high} = {\rm min}(z^j_{\rm max}, z^j_{K_{\rm lim}}, z^j_{i_{\rm lim}})
\end{equation}
\begin{equation}
z_{\rm low} = z^j_{\rm min},
\end{equation}

\noindent where the redshift interval, $j$, is defined by $[z^j_{\rm min},
z^j_{\rm max}]$ and $z^j_{K_{\rm lim}}$ and $z^j_{i_{\rm lim}}$  refer to the redshift at which the
galaxy would still be detected below the \kband\ and $i^+$-band limits.  We use the best-fit SED template as
determined by the stellar mass estimator to calculate these values,
thereby accounting for the $K$-corrections necessary to compute
$V_{\rm max}$ values (no evolutionary correction is applied).

\section{Results}\label{results}

\begin{figure*}[t]
\centerline{\includegraphics[scale=0.8]{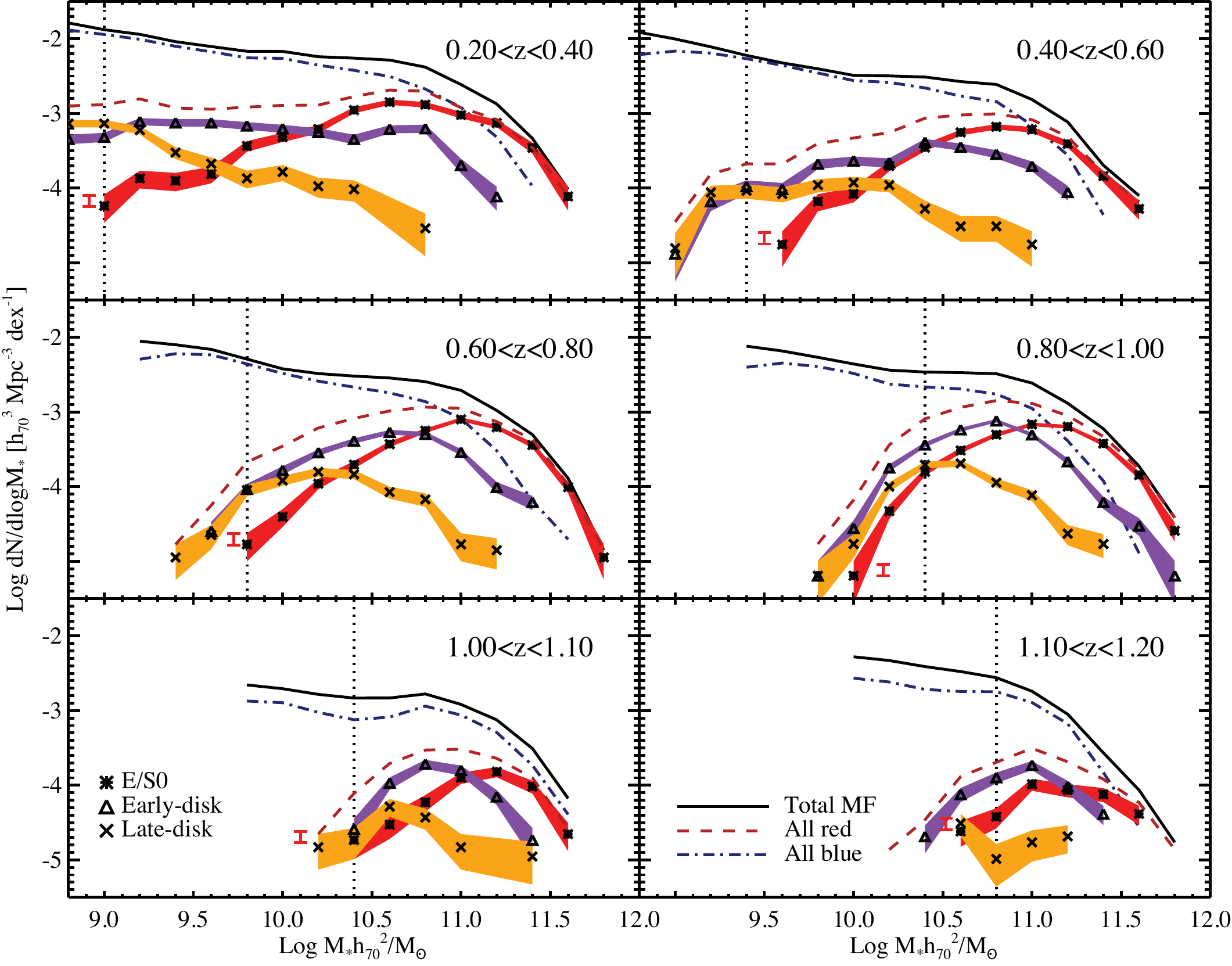}}
\caption{ In six redshift intervals, mass functions of galaxies on the red sequence divided by
  morphological type.  The total mass function is shown by the solid black line (the lowest redshift
  mass function is repeated as a gray line in all panels) while the abundance of passive galaxies
  (of all types) is shown as the dashed red line and that of star-formers with the blue dash-dotted
  line.  The passive (red-sequence) mass functions are further divided by morphological type as shown by the
  shaded regions which indicate the $1\sigma$ uncertainties.  Passive E/S0s have red shading with
  asterisks, passive early disks ($1.25 \leq \langle n \rangle < 2.5$) have purple shading with triangle symbols,
  and passive late disks ($0.75 \leq \langle n \rangle < 1.25$) have orange shading with X symbols.  The isolated
  error bars indicate the systematic uncertainty associated with a 30\% upward correction to the
  number of ZEST spheroidals potentially misclassified at the faintest magnitudes. The contributions
  from irregulars and passive galaxies with extreme disks (which are very rare) are not shown for
  clarity.  Cosmic variance tends to affect the abundances of sub-populations to a similar degree and
  introduces differences between redshift panels that can be as large as a factor of 4 but are
  typically closer to 60\%. \label{fig:mfn_abs}}
\end{figure*}

\subsection{Morphological Makeup of the Red Sequence}

\begin{figure}
\centerline{\includegraphics[scale=0.7]{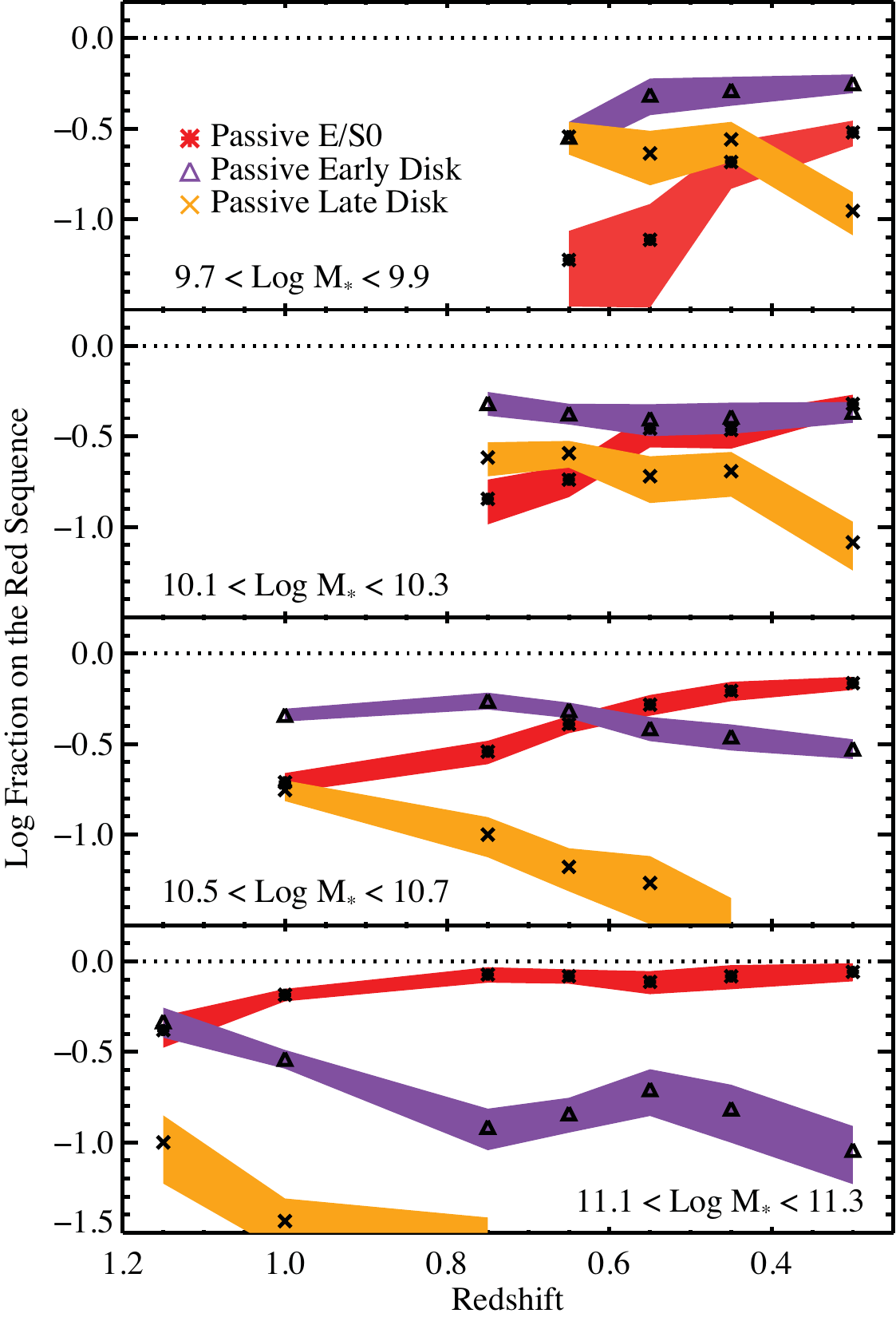}}
\caption{In 4 $M_*$ intervals, the fractional contribution to all red sequence (passive) galaxies
  from E/S0's, early-type disks, and late-type disks as a function of redshift.
  The contribution from irregulars is not shown for clarity.  Only data points above the 50\%
  completeness limit for obtaining successful morphological classification are
  plotted.  \label{fig:mfnf_time_morph}}
\end{figure}

Our main results are shown in Figures \ref{fig:mfn_abs} and \ref{fig:mfnf_time_morph} which present
the absolute mass function and fractional contribution of passive galaxies (which we refer to as the
``red sequence'') divided by ZEST morphological type into spheroidals, early disks, and late disks.
The mass distributions of all red-sequence galaxies (red dashed line), all star-forming galaxies
(blue dash-dotted line), and the full sample (black solid line) are also indicated.  The statistical
uncertainties are encoded by the width of the shaded regions.  The sample (or cosmic) variance in
COSMOS (see Section \ref{cosmic_variance}) is estimated at 0.2 dex and is likely responsible for the high
apparent abundance values (especially for passive objects) in the $0.8 < z < 1.0$ bin, a feature
seen in previous work \citep{ilbert10, drory09, pannella09, pozzetti09}.  Extreme disks with
passive colors are so rare that we do not plot them and for clarity we omit the morphological mass
functions of star-forming galaxies.  Though we do not show it, it is important to note that at low
masses star-forming disks of all three types far outnumber passive disks, while at the highest
masses they are roughly equal in absolute numbers.  The 50\% completeness of the ZEST morphological
sample is indicated with vertical dotted lines in Figure \ref{fig:mfn_abs} while data points below
this limit are simply omitted in Figure \ref{fig:mfnf_time_morph}.  These limits are nearly
identical to the 50\% completeness limits that would be obtained imposing a brightness limit of
$i_{814W} < 24$.

Figures \ref{fig:mfn_abs} and \ref{fig:mfnf_time_morph} show that galaxies on the red sequence, even
after removing those contaminated by dust, are remarkably diverse in their morphologies.  This may
have been missed in previous data sets that were limited to the highest masses where spheroidals do
indeed dominate the red sequence.  Thanks to the greater depth of the COSMOS photometry, the ZEST
morphologies reveal a significant population of red sequence galaxies at lower masses that appear to
harbor disk components.  These become prominent below a stellar mass scale that evolves downward
with time from 1--2$ \times 10^{11} \msun$ at $z \approx 1.2$ to $\approx$10$^{10} \msun$ at $z
\approx 0.3$.  Considering the full red sequence (defined by the mass range over which the fraction
of passive galaxies is $f_{\rm red} > 10$\%) we find that at every redshift, passive disks (early
$+$ late) are identified in as much as half of red-sequence galaxies and become dominant at the
low-mass end.  This increase in passive disks is robust to the maximum combined biases
that could result from morphological classifications as well as cosmic variance.  Dividing disks
further into ``early'' and ``late'' depending on whether the average S\'{e}rsic $\langle n \rangle$ values are greater than
or less than 1.25, we see that disk galaxies with prominent disk components (lower $\langle n \rangle$) are more
common at lower masses ($\lesssim$10$^{10}$ \Msun) becoming comparable to the more bulge-dominated
early disks with higher $\langle n \rangle$ values.  On the other hand, these early disks are more frequent at
higher masses and there is a redshift-dependent mass range (near $\sim$10$^{11} \msun$) where they
are roughly as abundant as passive spheroidals.  This general behavior seems to mirror the broader
pattern seen when the total population, regardless of rest-frame color, is split by morphological
type \citep[e.g.,][]{bundy05, ilbert09}.

The evolution of morphologies on the red sequence is most easily seen in Figure
\ref{fig:mfnf_time_morph} where the fraction of passive galaxies with a given morphology is plotted
as a function of redshift in four $M_*$ bins.  While spheroidals dominate the highest masses of the
red sequence at all times, at lower masses they start off as a small fraction of passive galaxies at
early epochs and rise in importance with time.  Meanwhile, the fraction of early and late passive
disks tends to decline with time.  As we discuss below, it seems plausible from Figure
\ref{fig:mfnf_time_morph} that the passive disks' fraction declines because they transform into
passive spheroidals, but Figure \ref{fig:mfn_abs} shows that their absolute numbers evolve less
strongly.  It is therefore likely that some star-forming systems (including star-forming
spheroidals) transform directly into passive spheroidals.  This is supported by Figure
\ref{fig:mfn_abs} which shows that at every redshift, there is an ample supply of star-forming
galaxies which may later transform into passive spheroidals.

\subsection{Verification of the Passive Disk Population}\label{verification}

\begin{figure*}
\centerline{\includegraphics[scale=0.8]{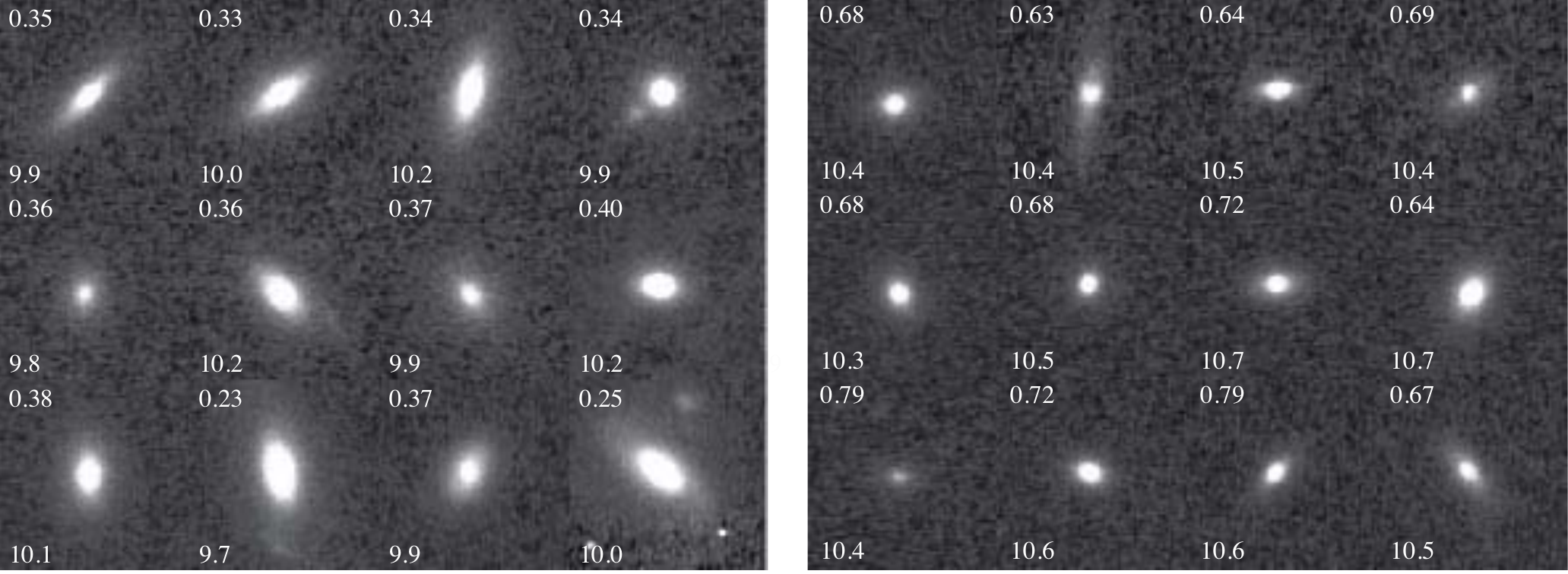}}
\caption{Examples of passive early disks identified using ZEST in the COSMOS data.  The
  left-hand panel shows galaxies with $0.2 < z < 0.4$ and $9.7 < \log M_*/\msun < 10.2$ while the
  right-hand panel corresponds to $0.6 < z < 0.8$ and $10.2 < \log M_*/\msun < 10.7$.  This class
  features bulge-dominated systems with average single-\Sersic\ $n$ values of $1.25 < \langle n \rangle < 2.5$.  The
  galaxies displayed were chosen at random from the subsample.  Their redshifts are indicated in the
  upper left of each postage stamp, the $M_*$ values in $\log \msun$ units in the lower left.  Each
  panel is 3\arcsec\ across, corresponding to roughly 12 kpc at $z \approx
  0.3$ and 21 kpc at $z \approx 0.7$. \label{fig:ex_RD_B1}}
\end{figure*}

\begin{figure*}
\centerline{\includegraphics[scale=0.8]{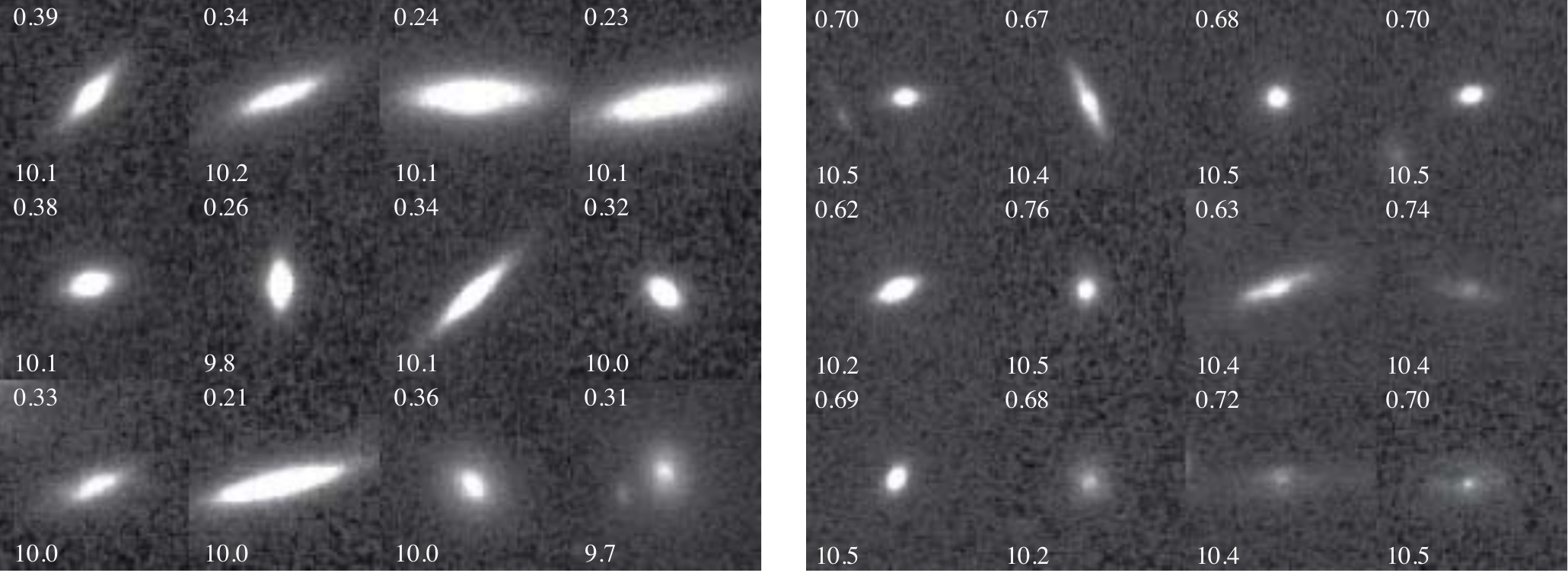}}
\caption{Examples of passive late disks with the two panels corresponding to the same mass and
  redshift range as in Figure \ref{fig:ex_RD_B1}.  These objects have average \Sersic\ $n$ values of
  $0.75 < \langle n \rangle < 1.25$ and are more disk-dominated than the early disks.  The postage
  stamp sizes and labeling is the same as in Figure \ref{fig:ex_RD_B1}.
  \label{fig:ex_RD_B2}}
\end{figure*}

Figures \ref{fig:mfn_abs} and \ref{fig:mfnf_time_morph} reveal a surprisingly abundant population
of passive galaxies that appear to contain disk components.  As we argue in Section \ref{discussion}, these
galaxies offer valuable clues to the mechanisms driving galaxy evolution.  We have visually
inspected early and late passive disks identified by ZEST in COSMOS and present random examples
drawn from specific redshift and $M_*$ ranges in Figures \ref{fig:ex_RD_B1} and \ref{fig:ex_RD_B2}.
Early disks tend to be bulge-dominated as expected, with many examples that might be classified as
Sa/S0 galaxies.  These galaxies often show rounded central regions surrounded by lower surface
brightness material that we attribute to the disk component responsible for the relatively low
\Sersic\ values ($1.25 < \langle n \rangle < 2.5$).  Note that our choice of ZEST classifications groups objects
exhibiting even higher $n$ values ($\langle n \rangle > 2.5$) in the spheroidals category.  The late disks (Figure
\ref{fig:ex_RD_B2}) tend to be more elongated with smaller bulges and comparably brighter disk
components.  In either category and across the full redshift range, there is little evidence for the
presence of significant amounts of dust and only a small degree of spiral structure.  Compared to
the clumpier appearance of star-forming disks with spiral arms and associated star forming regions,
the morphology of passive disk galaxies is consistent with the notion that star formation has been
largely shut down in these systems.

\begin{figure*}
\centerline{\includegraphics[scale=0.3]{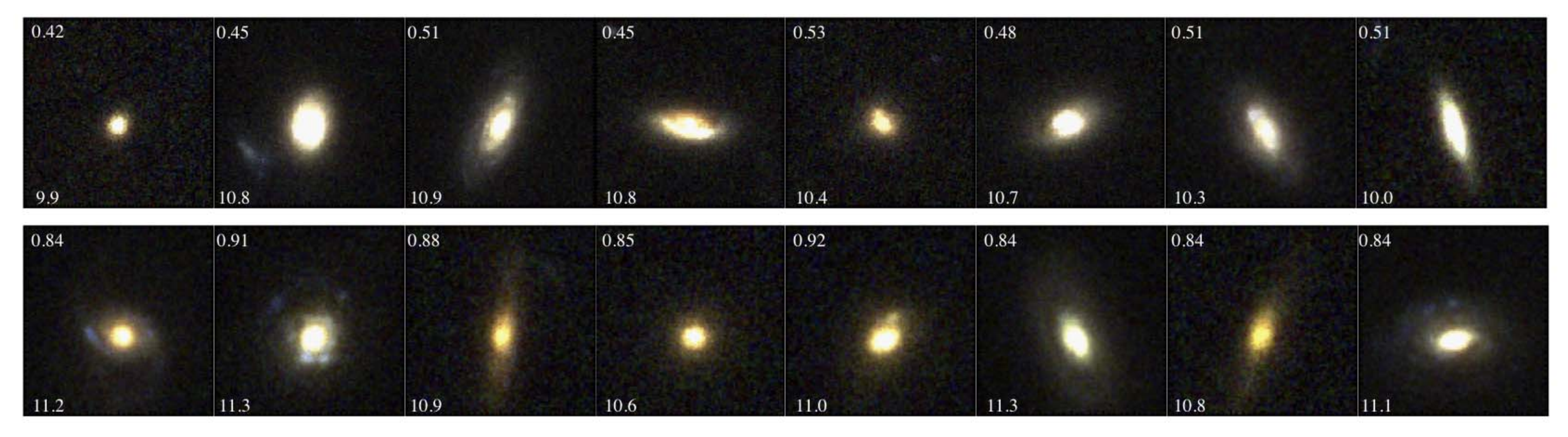}}
\caption{$Viz$ color images of passive galaxies in GOODS-N with {\em visually} classified disk
  morphologies.  The upper strip of eight postage stamps has $0.4 < z < 0.6$ while the lower panel corresponds
  to $0.8 < z < 1.0$.  As in Figure \ref{fig:ex_RD_B1}, the 
  width of each stamp is 3\arcsec\ and redshift is indicated in the upper left, $\log M_*/\msun$ in
  the lower left.   
  \label{fig:ex_RD_goods}}
\end{figure*}

While direct inspection supports the automated ZEST classifications, systematic biases could remain.
In Section \ref{morph}, we discuss how at faint magnitudes ZEST may misclassify up to 30\% of spheroidals
as disks.  This potential bias is illustrated by the systematic error bar in Figure
\ref{fig:mfn_abs} which shows that passive disks would still dominate the low-mass end, even if the
maximum 30\% correction were applied.  We have explored this bias further by comparing the
ellipticity distributions of passive and star-forming disks.  At the mass completeness limits of our
analysis, the ellipticities of passive and star-forming early disks are similar, but late passive
disks are weighted toward lower ellipticities.  This occurs at a level that is consistent with the
expected 30\% contamination of the late disk sample by misclassified early disks and spheroidals.
Thus, at the mass completeness limits and below, the passive late disk abundance should be viewed
with caution given that as much as 30\% could be misclassified.

We emphasize, however, that the basic trends in our results are robust because they are still
present when we limit our sample to bright magnitudes ($i_{814W} < 22.5$) where the classifications
are highly secure.  In absolute terms, we see by comparing to \citet{ilbert10} that other automated
methods classify some fraction of ZEST disks as spheroidals.  This is expected given that broad
divisions are somewhat arbitrary and that our goal is to identify evidence for disk components even
in bulge-dominated galaxies that might otherwise be grouped with spheroidals.  The galaxies with
disk morphologies in our sample are associated with average \Sersic\ indices of $\langle n \rangle <
2.5$ for the early disks and $\langle n \rangle < 1.25$ for late disks.  A value of $n=2.5$ has
often been used to divide disks from spheroidals \citep[e.g,][]{barden05}.  However, we should note
that surface brightness dimming at high redshifts may make some disk galaxies appear more
bulge-dominated, although this works in the opposite sense as the estimated bias at faint magnitudes
from \citet{scarlata07}.  So, until verification through bulge-to-disk decompositions is performed,
a conservative interpretation would be to view the abundance of ZEST disks as an upper limit.

A more specific issue is a possible bias from inclination.  We expect any type of disk component to
be more easily identified in edge-on systems where flatter isophotes are easier to distinguish.  Of
particular concern is the possibility that dust contamination in edge-on systems could hide residual
star formation given that the integrated colors could be dominated by a quiescent bulge.  To check
this, we have recomputed the mass functions of passive galaxies with the requirement that all
galaxies have a ZEST ellipticity class of 0, corresponding to completely face-on systems.  This
reduces the sample size by $\sim$80\%.  While the mass functions shift a bit under this assumption,
the essential behavior captured in Figure \ref{fig:mfnf_time_morph} is unchanged, with the same
relative abundance and mass dependence of the various morphological types preserved.  We can even
compute mass functions requiring significant elongation (ellipticity class 3).  While this disfavors
spheroidals because they are intrinsically rounder and boosts late disks which are intrinsically
flatter, the basic mass and redshift trends are still unchanged.  We can therefore rule out an
inclination bias in our ability to identify passive disks.

Another concern is the morphological $K$-correction which could mimic the rising abundance of
passive disks at early times if, as expected, galaxies appear more disk-dominated at bluer
rest-frame wavelengths.  Over the redshift range of the sample ($0.2 < z < 1.2$) the ACS $i_{814W}$
filter corresponds to a rest-frame wavelength range of 4000-6700 \AA.  It has been shown that
morphological $K$-corrections within this range are small for most galaxies \citep[e.g.,][]{lotz04,
  cassata05}, but with only one band of ACS data, it is hard to test this effect directly.
\citet{capak07} use a subsample of COSMOS galaxies to show that automated estimators are generally
robust to $K$-corrections, but to address this problem further and check our ZEST-based results more generally, we use the visually
classified morphologies in the northern GOODS-N \citep{giavalisco04} as presented in \citet{bundy05}.  These observations have a
comparable depth as the COSMOS $i_{814W}$ imaging, but include 4 {\it HST}/ACS bands ($BViz$), enabling
morphological classifications robust to potential $K$-corrections.

We identify quiescent stellar populations using similar color--color cuts for the GOODS data as for
COSMOS and do indeed find a population of passive disks with {\em visual} morphological types of
Sa--Scd.  Figure \ref{fig:ex_RD_goods} shows several examples.  While the small size of the GOODS
comparison sample prevents a detailed comparison, it nonetheless shows similar evidence for an
increasing fraction of passive disks as a function of redshift.  For $0.4 < z < 0.7$ and above the
completeness limit of $\log M_*/\msun \approx 10.2$ we find $\lesssim$20\% of the quiescent
population is identified as disk-like.  At $1.0 < z < 1.1$ and $\log M_*/\msun \gtrsim 10.7$, the
fraction rises to roughly 35\%.  While the statistical uncertainty is large, the sense of this trend
reinforces the results of the larger COSMOS sample.  It should be noted that biases among visual
morphologies favor early type classifications at faint magnitudes and high redshifts due to surface brightness
dimming \citep[e.g.,][]{brinchmann98}.  We also note that the GOODS objects show no apparent bias
with inclination, in agreement with the inclination tests above.  The use of color information in
refining the GOODS morphologies indicates that the evolutionary signal is not likely to result from
$K$-corrections and the fact that passive disks identified visually exhibit similar abundances and
evolution lends support to the ZEST classifications.

\section{Discussion }\label{discussion}

We have used automated $i_{814W}$-band ACS morphologies in the COSMOS survey supported by visual
morphological classifications of $BViz$ ACS imaging in GOODS-N to argue for the existence of a
significant population of passive, bulge-dominated galaxies hosting disk components.  These galaxies
were more abundant in the past ($z \sim 1$) and are seen to be more common among galaxies with lower $M_*$
at all redshifts probed.  In fact, at the low-mass end they represent as much as 70\%--90\% of the red
sequence, even after dusty star-forming red galaxies are removed.  At these masses, passive disks
account for about 10\% of all galaxies.  The fact that their abundance depends on both $M_*$ and
redshift, and that these dependences are influenced by the strength of the disk component argues
that passive disks hold clues to understanding both the quenching of star formation and
morphological evolution within the galaxy population more broadly.  We begin our discussion with an
assessment of how important passive disks are to galaxy evolution, finding that as many as 60\% of
galaxies transitioning onto the red sequence may evolve through a passive disk phase.  We then
discuss possible formation mechanisms, first ruling out simple fading of star-forming disks and
strong environmental processes and then moving on to discuss disks regrown in mergers and the
suggestion that quenching is initiated by internal structural changes.  While no current model fully
reproduces the observations at present, our constraints will help refine proposed quenching
mechanisms and indicate that morphological and color transformations in galaxies proceed in multiple
and sometimes separate stages.

\subsection{The Importance of Passive Disks as a Phase of Evolution}\label{importance}

How significant are passive disks in the evolutionary history of galaxies?  Ideally, we could answer
this question by tracking the mass-dependent flow of galaxies transitioning through different
evolutionary phases, but unfortunately, as discussed in Section \ref{cosmic_variance}, the strong cosmic
variance uncertainties in the single COSMOS field prevent a detailed treatment.  Broadly speaking,
Figures \ref{fig:mfn_abs} and \ref{fig:mfnf_time_morph} in conjunction with previous work suggest
that passive disks may represent a significant and relatively common phase of evolution.  While the
red sequence abundance grows with time \citep[e.g.,][]{bundy06, borch06}, passive disks always
dominate at lower masses which is also where the newest arrivals to the red sequence can be found
\citep[e.g.,][]{treu06}.  This suggests that a large fraction of these quenching galaxies experience
an initial passive disk phase.  This is supported by Figure \ref{fig:mfn_abs} which shows that, as
with red-sequence spheroidals, the absolute number density of passive disks below $\sim$10$^{11}
\msun$ {\em increases} with time, even though, compared to all passive galaxies in this mass range,
their fraction {\em decreases} with time.  Choosing a mass between 10$^{10}$ and 10$^{11} \msun$
and comparing the highest and lowest redshifts sampled, we see that $\sim$20\%--40\% of new red
sequence galaxies formed over this interval can be accounted for by the increase in the number of
passive disks.  At masses above $\sim$10$^{11} \msun$, evolution in the absolute abundance of
passive disks is more difficult to discern but, as with their fraction, likely declines by 20\%--50\%
(compare the two lower redshift bins in Figure \ref{fig:mfn_abs} and note the lack of high-mass
examples at late times).  This reinforces the visual impression of Figure \ref{fig:mfnf_time_morph},
namely that passive disks evolve into spheroidals.  Indeed, as we discuss below, one expects
morphological evolution of passive disks to be enhanced compared to their star-forming counterparts
because lower gas fractions ($f_{\rm gas}$) enable more minor and, hence, more frequent mergers to
drive transformations \citep{hopkins09}.  If we make the assumption that all passive disks transform
into spheroidals within 1--2 Gyr \citep[e.g.,][]{van-dokkum05}, the fraction of new arrivals to the
red sequence transitioning through a disk phase must increase to compensate, and could be as high as
60\%\footnote{For a mass between 10$^{10}$ and 10$^{11} \msun$ this percentage can be estimated by
  assuming that all of the passive disks at $z \approx 1$ become passive spheroidals by $z \approx
  0.7$ (an interval of roughly 1.5 Gyr).  The remaining spheroidals that have formed between these
  two redshifts therefore originate from star-forming galaxies.  This accounts for $\sim$40\% of the
  total growth of the red sequence abundance over this interval.  Thus, $\sim$60\% of red sequence
  growth must take the form of passive disks.}.  Passive disk galaxies may therefore represent a
fairly common phenomenon and highlight the importance of further morphological evolution along the
red sequence.

\subsection{Can Passive Disks Form From Faded Star-forming Disks?}\label{concentration}

\begin{figure}[t]
\centerline{\includegraphics[scale=0.5]{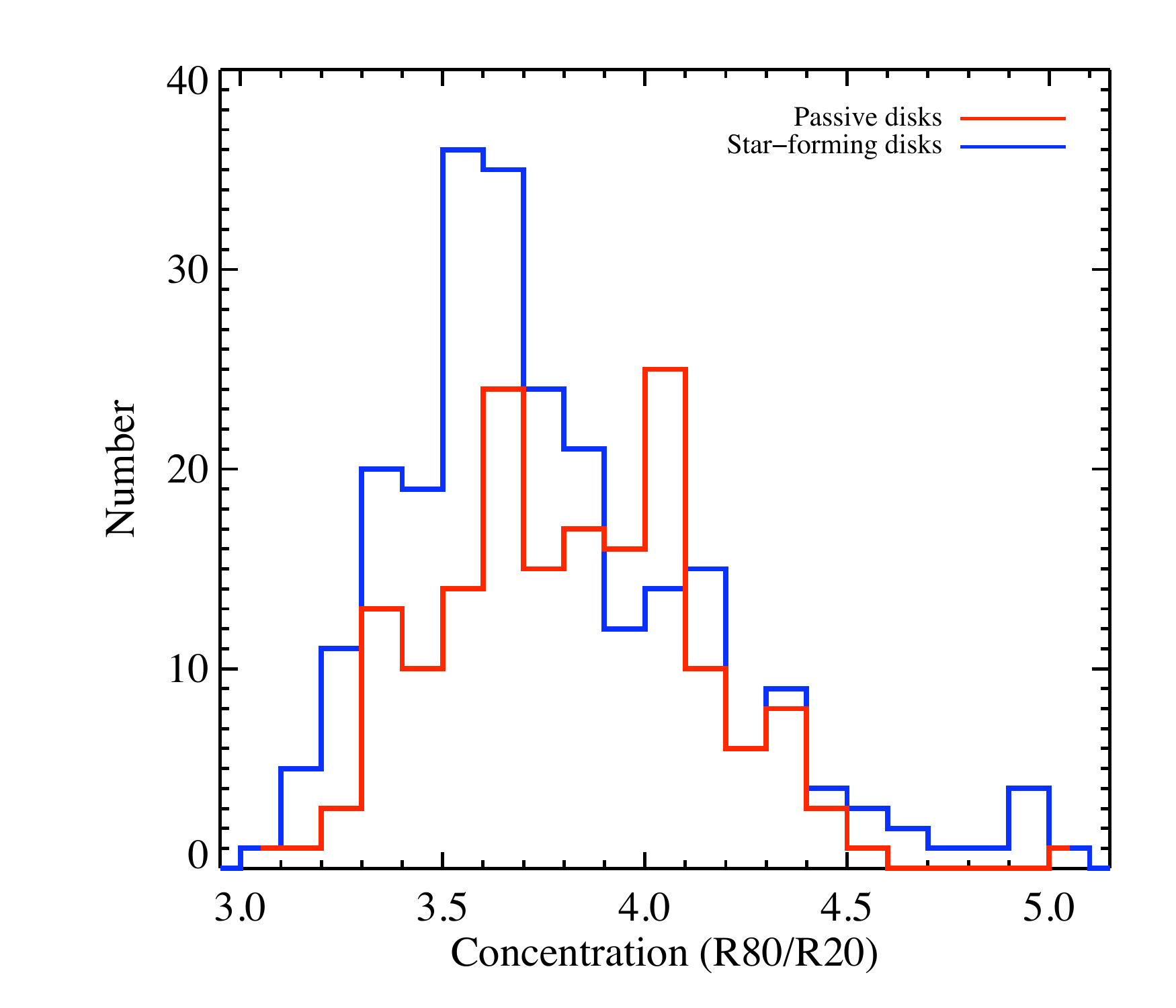}}
\caption{ Concentration of passive vs.~blue disks of comparable morphologies in the $0.2 < z <
  0.4$ redshift range with $9.7 < \log M/\msun < 10.2$.  Passive disks in the same morphological
  class show evidence of higher concentrations.  The ACS $i_{814W}$ band samples the restframe at
  $\sim$6200\AA, thus avoiding potential classification biases owing to star-forming
  regions.  \label{fig:disk_C}}
\end{figure}

At all times, there are as many if not more star-forming disks than passive disks, so it is easy to
imagine that passive disks may be drawn from the star-forming population with one important caveat:
there are almost no passive galaxies with ``extreme disk'' (\Sersic\ $\langle n \rangle < 0.75$)
morphologies.  Such extreme disks are common in large numbers among star-forming galaxies at all
masses and even dominate at the lowest masses where passive disks are most abundant.  They would be
easily detected if they existed in the passive population.  Therefore, the transformation to the red
sequence either leads to the formation a bulge component \citep[e.g.,][]{dominguez-palmero09} or
requires the existence of one.  In the latter case, star-forming and passive disks with similar
bulge components should differ only in color but not in their intrinsic structure.  We will postpone
a detailed comparison of bulge/disk decompositions, scale lengths, and $M/L$ evolution to future
work and focus here on a simple comparison of the concentration ($C=R_{80\%}/R_{20\%}$) of passive
and star-forming early disks (Figure \ref{fig:disk_C}).  We consider the nearest redshift bin $0.2 <
z < 0.4$ so that the ACS $i_{814W}$-band probes the reddest possible wavelengths (least affected by
star formation), and consider the mass range of $9.7 < \log M_*/\msun < 10.2$.  Passive disks appear
to be moderately more concentrated with a peak at $C \approx 3.9$ compared to blue disks with a peak
of $C \approx 3.6$ (a Kolmogorov-Smirnov probability test shows that the probability of 0.02 for the distributions
being drawn from the same sample).  Even after dividing the star-forming disks into low and high SFR
subsamples (based on their NUV luminosities) and comparing just the low SFR star-formers to passive
disks, this difference in concentration remains.  It also holds to $z \approx 0.9$ despite
rest-frame morphologies that sample bluer wavelengths.  Still, this signature could largely be
driven by differences in $M/L$ gradients between star-forming and passive systems.  Given this
tentative result and the stronger requirement that passive disks be more bulge-dominated than
star-forming disks, it seems unlikely that passive disks form simply via the quenching and fading of
star formation disk galaxies.

\subsection{The Role of Environment}\label{environment}

Could passive disks be a product of dense environments?  Their morphologies are sometimes
reminiscent of S0 galaxies (see Figure \ref{fig:ex_RD_goods}) which are thought to become
increasingly common in clusters \citep[e.g.,][]{dressler97, smith05} although recent work disputes
this \citep{holden09}.  While we have specifically attempted to exclude S0s from our disk sample
(they are classified as spheroidals), distinguishing them at distant epochs remains difficult
\citep[e.g.,][]{moran07}.  A more promising link exists with ``passive spirals'' (defined via far UV
and spectra line diagnostics) which likely transform into S0s in cluster environments
\citep{moran05, moran06, moran07a}. This transformation process is thought to begin slowly (1--2 Gyr
timescales) in infalling groups, perhaps as a result of ``gentle'' interactions with other galaxies
and the intergroup medium.  If the intracluster medium (ICM) is sufficiently dense, ram pressure stripping strongly
accelerates the quenching of star formation, producing S0s and ellipticals closer to the cluster
core and driving evolution in the morphology--density relation \citep[e.g.,][]{capak07}.

Because there are only 1--2 clusters in the 1.6 deg$^2$ COSMOS field \citep{meneux09}, cluster
mechanisms cannot fully account for the passive disk population studied here.  Still, group scale
environments may play an important role \citep[see][]{wilman09}.  To investigate further, we define
a ``field'' sample by removing sources along the line-of-sight to dense structures.  These were
identified in the X-ray-selected COSMOS group catalog assembled by \citet{finoguenov07a} and
\citet{leauthaud10} which contains 206 groups with weak-lensing calibrated virial (total) masses
complete to $M_{200} \approx 3 \times 10^{13} \msun$ to $z = 0.5$ and $M_{200} \approx 3 \times
10^{14} \msun$ to $z = 1.0$.  Any galaxy within the projected $R_{200}$ virial radius of any group,
regardless of its redshift, is removed from the field sample, reducing the full sample by
$\approx$30\%.  Remarkably, the mass dependence and evolution of passive galaxy morphologies (Figure
\ref{fig:mfnf_time_morph}) in this field sample is virtually identical to the full sample.  This is
not to say that there are no environmental trends in the passive disk fraction which would be
expected from previous studies of morphology--density relation.  Our ``field'' test simply shows that a large fraction of
passive disks live and evolve in halos less massive than $10^{13} \msun$ (given our two
lowest $z$-bins), ruling out environmental processes in more massive halos as the {\em sole}
explanation.

Even at $M_{200} \approx 10^{13} \msun$ environmental effects can be important as very weak ram
pressure (2--3 orders of magnitude weaker than in cluster cores) can cause ``strangulation'' or
``starvation'' via stripping of the galaxy's hot gas halo, thereby shutting off a potential fuel
source for subsequent star formation \citep[see][]{mccarthy08, bekki09}.  The starvation scenario
would help explain the prevalence of passive disks at low masses, since simulations show that lower
mass galaxies are more easily affected.  But the starved remnants should have structural properties
similar to star-forming disks, a result that seems to be at odds with Figure \ref{fig:disk_C}.  In
addition, \citet{mccarthy08} show that less concentrated disks are more susceptible to starvation.
Yet, the very low-concentration ``extreme disk'' morphologies found in the star-forming population
are almost completely absent among passive galaxies.  Furthermore, because as $M_*$ increases
starvation becomes less effective, we would expect the least concentrated passive disks (the ``late
types'') to occur at the highest masses.  Our observations show the opposite trend.  We also might
expect the group abundance and typical gas density, $n_{\rm ICM}$, to grow with time from $z=1.5$
\citep[e.g.,][]{mcgee09}.  Yet, the most massive examples of passive disks, for which $n_{\rm ICM}$
must be large to induce starvation, occur in our {\em highest} redshift bins ($z \sim 1$).

We therefore conclude that while it is possible that passive disks are a product of low-mass group
environments, proposed environmental processes at these scales appear to have some difficulty
explaining the observed mass dependence, evolution, and concentration of these galaxies.

\subsection{Gas-rich Mergers, Disk Regrowth, and AGN Quenching}

Beyond environmental processes, star formation quenching has often been associated with feedback
mechanisms triggered by major mergers \citep[e.g.,][]{hopkins08}.  The existence of passive disks
presents a problem for this hypothesis: how can a major merger shut down star formation in these
galaxies without destroying the underlying stellar disk?  One possible solution is that before
quenching completes, an entirely new disk is formed \citep{hammer05}.  Recent high resolution
merger simulations have illustrated exactly this effect \citep{springel05c, robertson06,
  governato09}.  Roughly speaking, new gaseous disks can form in gas-rich mergers with $f_{\rm gas}
> 0.5$ because a lack of central instabilities prevents gas transport to the center and suppresses
bulge growth.  Given observations of $f_{\rm gas}$ as a function of $M_*$, mergers with $M_* \approx
10^{10} \msun$ are much more likely to result in a remnant with a disk-like morphology than similar
mergers at $M_* \approx 10^{11} \msun$ \citep[see][]{hopkins09}.

\begin{figure}[t]
\centerline{\includegraphics[scale=0.55]{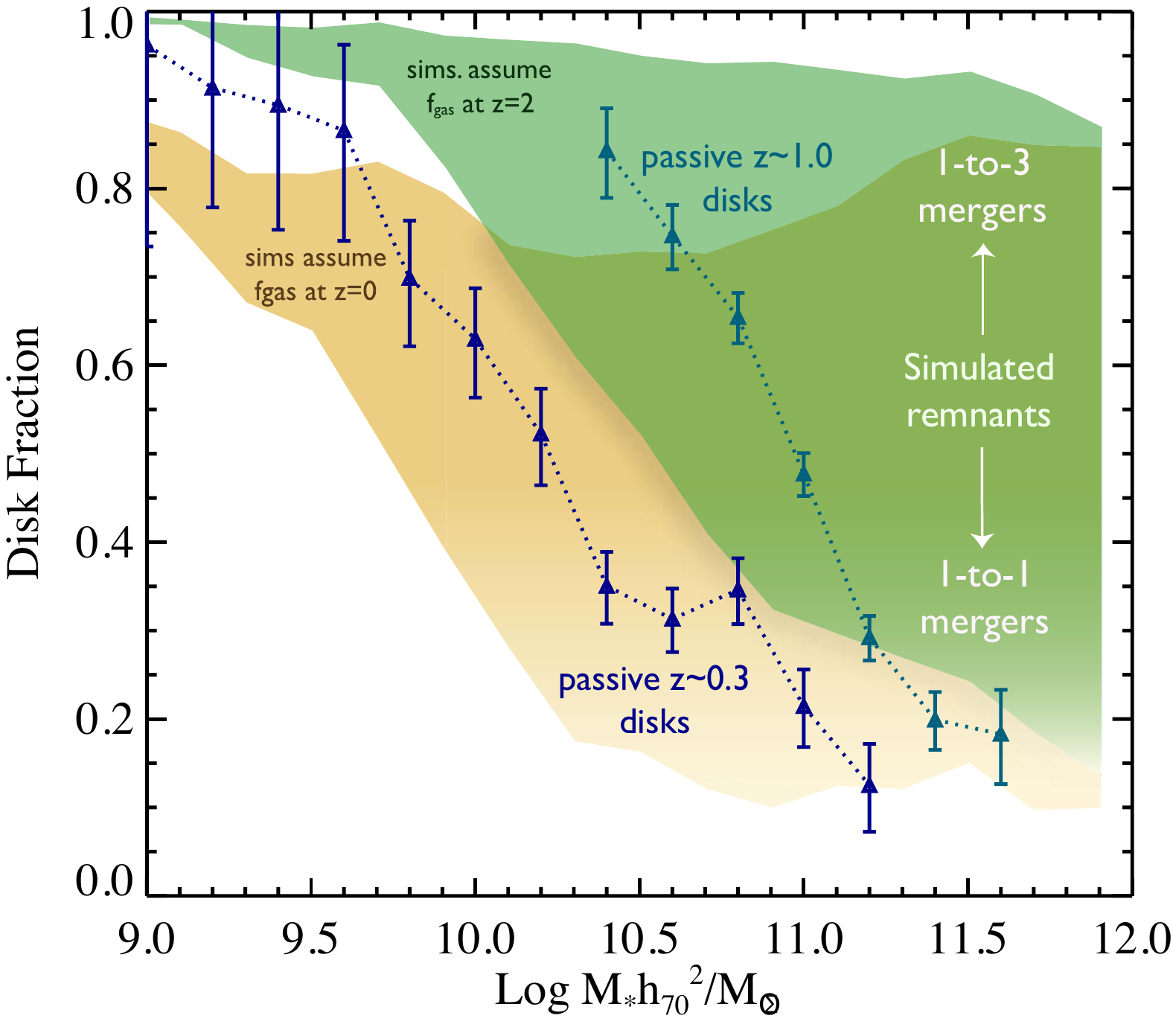}}
\caption{Comparison of the predicted fraction of simulated merger remnants that would be classified
  as disks to the observed fraction of red-sequence galaxies with disk morphologies (early $+$ late
  types).  The shaded regions denote predictions for two redshift ranges, $z \sim 1$ and $z \sim
  0.3$.  In either case, low mass ratio mergers predict a high abundance of remnant disks while
  one-to-one mergers trace the decreasing amount $f_{\rm gas}$ as a function of $M_*$.  Overplotted
  are the observed passive disk fractions in the two $z$-bins.  The toy model matches the observed
  passive disk fractions, as well as their mass and redshift dependence, if most formed in mergers
  with $\mu \sim 1.0$.  \label{fig:gas_disk_remnants}}
\end{figure}

One of the key aspects of the merger/regrowth model is that the strength of the remnant disk
component depends on $f_{\rm gas}$ \citep{stewart09a, hopkins09c}.  To investigate how predictions
from this model compare to our observations, we used the relationships between $f_{\rm gas}$, merger
mass ratio ($\mu$), $M_*$, and remnant bulge-to-total ratio ($B/T$) that \citet{hopkins09} derive
based on a large suite of merger simulations.  We assume a scatter of 0.25 dex in the predicted
$B/T$ and consider mergers with $0.3 < \mu < 1.0$.  We use $f_{\rm gas}(M_*)$ evaluated at $z = 2$
to correspond to passive disk progenitors at $z \sim 1$, and assume that the $z=0$ relation applies
adequately to progenitors at $z \sim 0.3$.  Remnant disks are those with $B/T < 0.6$.  The results
of this exercise are plotted in Figure \ref{fig:gas_disk_remnants} as two shaded regions for the two
redshifts evaluated.  These regions represent the predicted disk fractions spanned by mergers of
different mass ratios.  The region corresponding to the $z=0$ $f_{\rm gas}(M_*)$ is offset to lower
$M_*$, reflecting the decline of $f_{\rm gas}$ at higher masses.  Overplotted are the observed
fractions of passive disks (early $+$ late disks) at the two redshifts considered.  While in detail,
the scatter in predicted $B/T$ likely depends on mass ratio and $f_{\rm gas}$, this toy remnant
model does a good job matching the observed passive disk fractions as well as their mass and
redshift dependence if most formed in mergers with $\mu \sim 1.0$.  A more detailed comparison
between predicted and observed morphologies is needed to test this further.

This merger/regrowth scenario is consistent with the lack of a strong environmental dependence
(Section \ref{environment}) and would predict that some passive disks will be isolated.  This could be
tested through clustering and pair count studies.  Furthermore, the somewhat higher concentration in
passive disks compared to star-formers is naturally explained by the extra bulge mass added during
the merger.  The big drawback remains understanding how mergers lead to the quenching of star
formation.  If quenching is short-lived and violent, as in models employing a quasar phase
\citep[e.g.,][]{menci06, hopkins08}, it must somehow be delayed until enough time passes for a
stellar disk to regrow (at least 1 Gyr).  At the same time, recent simulations have cast doubt on
the ability of quasar-mode AGN feedback to directly quench star formation \citep[e.g.,][]{debuhr10a}.

Instead AGN may be responsible for subtle feedback cycles that prevent gas cooling on longer
timescales (1--2 Gyr) after the cold gas is depleted \citep[e.g.,][]{croton06}.  If this feedback is
tied to the central black hole mass, this could explain why passive disks have large bulges
(Section \ref{concentration}), but it does not make clear why other star-forming disk galaxies with
similar bulges have not been quenched (Figure \ref{fig:disk_C}).  Finally, if quenching is initiated
by low luminosity AGN feedback, are mergers even required \citep[e.g.,][]{ciotti07}?  While the
specific quenching process in these scenarios remains unknown, the agreement in Figure
\ref{fig:gas_disk_remnants} nevertheless supports a formation mechanism in which the frequency and
appearance of passive disks is strongly linked to their progenitor gas fractions.

\subsection{A Two Stage Scenario for Morphological Evolution}

If star formation in passive disks is shut down because the gas supply is depleted, simulations show
that such gas-poor disks can be more easily disrupted by mergers with lower mass ratios---that is,
more minor interactions---than disks with high $f_{\rm gas}$, owing to the ability of gas to regrow
disks and suppress bulge formation \citep{hopkins09}.  This suggests a two-step process where first,
late-type star-forming disks are quenched and transformed into bulge-dominated, passive disks.  This
makes them vulnerable to minor interactions that, second, disrupt the disk component and convert
them into spheroidals\footnote{We avoid the term ``dry mergers'' here, even though the interactions
  considered would have low $f_{\rm gas}$, because dry mergers have come to denote the more narrow
  concept of nearly binary mergers between gas-poor, but already spheroidal galaxies.}.

Figure \ref{fig:mfn_abs} suggests that this two-step evolution may have a broad impact on
interpretations of how morphological populations evolve.  The figure shows that passive disks
account for as much as one-third of all disk galaxies at $M_* \sim 10^{11} \msun$ while the number
declines toward lower masses (and redshifts).  Assuming they are gas-poor and susceptible to more minor
interactions, then because minor mergers are far more common than major mergers
\citep[e.g.,][]{stewart09}, this passive one-third of the disk population is likely to account for
many of the newly formed spheroidals since $z \sim 1$ \citep[e.g.,][]{bundy05, ilbert10}.  But
recent efforts to compare the formation rate of spheroidals to the merger rate have focused only on
rarer, {\em major} mergers, assuming perhaps incorrectly that single mergers of gas-rich disks are
the only channel for spheroidal formation \citep[e.g.,][]{bundy07, genel08, bundy09}.  These studies
have suggested that the major merger rate, both as observed and predicted, may be 2--3 times too low
to explain the formation of new spheroidals.  While a more precise comparison requires consistency
in defining mass ratios, morphologies, and understanding gas effects \citep{hopkins10} 
including more prevalent minor mergers involving passive disks could help bring current measures
of the two phenomena into better agreement.

\subsection{Morphological Quenching}

The environmental effects and merger/regrowth model we have considered depend on phenomena that are
external to the galaxy.  It is also possible that passive disks form as a result of {\em internal}
processes.  An example of this is ``morphological quenching'' \citep[MQ][]{martig09}.  In this
model, after the cold streams that fuel high-$z$ star formation shut down or become clumpy
\citep{dekel09}, the further fragmentation of gaseous disks---and therefore its ability to form
stars---can be internally suppressed by two processes.  The first is the declining self-gravity of
the gas as it is converted to stars and the second is the deepening central potential caused by
stellar bulge growth resulting from instabilities in the stellar disk or even minor mergers.  The
predicted remnants of morphological quenching are consistent with many of our observations.  They
would appear as passive disks with large bulges, would have a weak environmental dependence, and
exhibit slow quenching, in this case initiated by gradual changes in internal structure.

While more detailed comparisons between observations and simulations of morphological quenching
\citep{martig09} are clearly needed, we note that many of the examples in Figures
\ref{fig:ex_RD_B1}--\ref{fig:ex_RD_goods} show large, thick disks.  In future work, this may help
distinguish between an early formation with thickening from disk instabilities that could accelerate
MQ \citep{bournaud09} and later disk regrowth \citep{hopkins09}.  It is also important to determine
if MQ can explain why bulge growth leads to quenching in some galaxies but not in others as well as
why passive disks are more disk-dominated at lower $M_*$ (Figure \ref{fig:mfn_abs}).  The higher gas
fractions at these masses would naively require larger, not smaller bulges.  Finally, we note that
morphological quenching requires testable deviations from the Kennicutt-Schmidt relation
\citep[e.g.,][]{donovan09} and implies that gas disks in passive galaxies remain intact.  Subsequent
mergers of passive disks in this scenario could therefore produce new star-forming galaxies instead
of passive spheroidals.

\subsection{Comparison with an SDSS Sample}

Finally, while this work has focused on mass-limited samples at high redshift, a detailed study by
\citet{masters10} of a subset of passive disks identified using GZ in SDSS provides
additional insight.  A direct comparison is not possible because the GZ passive disks were selected
using a single color ($g - r$) and were required to be mainly face-on and to have spiral arms as
identified by GZ users.  Still, the two samples paint a consistent picture.  Red GZ disks
are more common at higher masses \citep[see also][]{van-der-wel09} and tend to host larger bulges
than their star-forming counterparts.  Their emission lines indicate little ongoing SF but are
inconsistent with post-starburst spectra, favoring instead a gradual decline of SF over 1--2 Gyr.
They tend to live in intermediate environments consistent with the $\lesssim$10$^{13} \msun$ halos our
observations suggest.  More than 75\% host nuclear bars and 50\%--80\% show evidence for LINER
emission.  Assuming the GZ sample is representative of passive disks in general, the
\citet{masters10} analysis supports a formation mechanism that is strongly tied to a bulge and a
central bar, somewhat weakly tied to environment, and capable of quenching star-formation slowly.

\section{Summary and Conclusions}\label{summary}

We have used the COSMOS survey to identify a significant population of non-star-forming, passive
disk galaxies and have studied their importance with respect to the growth of the red sequence and
the formation of spheroidal galaxies.  After removing dusty contaminants, evidence for disk
components is found in nearly half of galaxies on the red sequence at all redshifts, primarily at the low-mass
end.  The absolute number of passive disks below $\sim$10$^{11} \msun$ appears to increase with
time, mirroring the growth of the red sequence, although the number of passive spheroidals increases
faster.  Above $\sim$10$^{11} \msun$, the abundance of passive disks has declined since $z \sim 1$.
We interpret these trends to indicate that passive disks may be a relatively common phase of galaxies
transitioning onto the red sequence, accounting for as much as 60\% of newly quenched systems.  The
declining contribution of passive disks to the red sequence with time suggests that, once formed,
they transform into spheroidals on moderately fast (1--3 Gyr) timescales.  This is likely aided by the
fact that unlike star-forming galaxies, passive disks may be more susceptible to morphological
evolution induced by minor mergers, given their depleted gas supply \citep{hopkins09c}.  Thus the
formation of many red sequence ellipticals does not occur in a single event like a major merger, and
may instead proceed through two or more evolutionary stages.

Understanding how passive disks form can shed light on the physical processes that shape galaxy
evolution more broadly.  Do passive disks evolve slowly from star-forming disks that exhaust their
gas supply after cooling is suppressed or become stabilized to fragmentation?  Or, do they form
more rapidly after some event, such as a merger or falling into a galaxy group?  While more work is
needed, early results from this analysis and a study of a subset of passive disks in SDSS
\citep{masters10} provide tantalizing clues.  First, passive disks are almost exclusively
bulge-dominated.  Star-forming disks on the other hand include a large number of strongly
disk-dominated galaxies at all epochs.  A bulge component is either required to shut down star
formation---perhaps because a bulge signals the presence of a potential AGN---or is built during the
formation process.  The presence of a large bulge, however, does not guarantee that star formation
is shut down.  Beyond this, passive disks with the same mass, redshift, and morphological type are
more concentrated on average than star-formers.  While these assertions should be verified with stellar mass
bulge/disk decompositions, it appears that at least some fraction of passive disks is structurally
distinct.  

We have used a toy model to show that mergers, which naturally increase bulge mass but can also form
a new disk \citep[e.g.,][]{springel05c}, produce remnants with a similar mass and redshift
dependence as passive disks.  This agreement suggests a strong link between morphology and $f_{\rm
  gas}$.  The problem with the merger/regrowth model remains understanding how the merger induces
quenching.  SDSS observations of passive disk emission lines favor a slow process \citep{masters10}
that operates over 1--2 Gyr.  This seems inconsistent with a merger-triggered quasar phase, which
must be delayed to allow enough time for a new disk to form.  On the other hand, slow quenching may
take place in 10$^{13} \msun$ halos in which the group medium can slowly ``starve'' galaxies of fuel
by stripping their hot gas reservoirs.  Environmental processes therefore present an alternative to
the merger/regrowth model but have trouble accounting for the importance of bulges and the way in
which passive disk morphologies depend on mass and redshift.  The newly proposed morphological
quenching scenario \citep{martig09} also predicts slow quenching as internal structural changes
suppress disk fragmentation and star formation.  This mechanism requires large bulges, as observed,
and predicts thick stellar disks \citep{bournaud09}.  More detailed comparisons and a test of the
predicted mass and redshift dependences in this scenario are now needed.  

The prospects for resolving the formation mechanism look promising and we note that some combination
of the processes considered here may be operating simultaneously.  More detailed studies of the
structure, color gradients, and stellar populations of passive disks will help link them with their
progenitors.  At the same time, it is important to verify the high disk fractions revealed by the
ZEST classifier with bulge-to-disk decompositions, ideally performed in terms of stellar mass.  This
will confirm the significance of disk components and enable more direct comparisons to star-forming
samples as well as simulations.  Improvements at low-$z$ will come from expanded, mass-limited
samples, and at high-$z$ from clustering and pair-count studies.  Larger samples from future surveys
will overcome cosmic variance and allow individual populations to be tracked as they evolve through
different stages.  Both with current surveys and in the future, passive disks offer a unique
opportunity to separate and understand the processes that drive global patterns of galaxy evolution
since $z \sim 1$.

\section{Acknowledgments}

We thank Eliot Quataert, Arjen van der Wel, Avishai Dekel, Fr\'{e}d\'{e}ric Bournaud, Bob Nichol,
Karen Masters, and Tommaso Treu for very useful discussion and feedback.  KB acknowledges support
for this work provided by NASA through Hubble Fellowship grant \#HF-01215, awarded by the Space
Telescope Science Institute, which is operated by the Association of Universities for Research in
Astronomy, Inc., for NASA, under contract NAS 5-26555.  We acknowledge the entire COSMOS
collaboration which has made this work possible.  More information on the COSMOS survey is available
at {\tt http://www.astro.caltech.edu/cosmos}.


\end{document}